\documentclass[aps,preprint,amssymb]{revtex4}

\usepackage{epsfig}

\begin{document}
\renewcommand{\thefootnote}{\fnsymbol{footnote}}
\renewcommand{\theequation}{\arabic{section}.\arabic{equation}}

\title{Specific Na$^{+}$ and K$^{+}$ Cation Effects on the Interfacial Water Molecules at the
Air/Aqueous Salt Solution Interfaces Probed with Non-resonant Second
Harmonic Generation (SHG)}

\author{Hong-tao Bian\footnote[2]{Also Graduate University of the Chinese Academy of
Sciences.}, Ran-ran Feng\footnotemark[2], Yuan Guo and Hong-fei
Wang\footnote[1]{To whom correspondence should be addressed. E-mail:
hongfei@iccas.ac.cn; Tel: 86-10-62555347; Fax:86-10-62563167.}}


\affiliation{Beijing National Laboratory for Molecular Sciences,
State Key Laboratory of Molecular Reaction Dynamics, Institute of
Chemistry, Chinese Academy of Sciences, Beijing, China 100190}

\date{\today}

\begin{abstract}
\noindent Here we report the polarization dependent non-resonant
second harmonic generation (SHG) measurement of the interfacial
water molecules at the aqueous solution of the following salts: NaF,
NaCl, NaBr, KF, KCl, and KBr. Through quantitative polarization
analysis of the SHG data, the orientational parameter D,(D=$\langle
cos\theta\rangle/\langle cos^3\theta\rangle$) value and the relative
surface density of the interfacial water molecules at these aqueous
solution surfaces were determined. From these results we found that
addition of each of the six salts caused increase of the thickness
of the interfacial water layer at the surfaces to a certain extent.
Noticeably, both the cations and the anions contributed to the
changes, and the abilities to increase the thickness of the
interfacial water layer were in the following order: KBr $>$ NaBr
$>$ KCl $>$ NaCl $\sim$ NaF $>$ KF. Since these changes can not be
factorized into individual anion and cation contributions, there are
possible ion pairing or association effects, especially for the NaF
case. We also found that the orientational parameter D values of the
interfacial water molecules changed to opposite directions for the
aqueous solutions of the three sodium salts versus the aqueous
solutions of the three potassium salts. These findings clearly
indicated unexpected specific Na$^{+}$ and K$^{+}$ cation effects at
the aqueous solution surface. These effects were not anticipated
from the recent molecular dynamics (MD) simulation results, which
concluded that the Na$^{+}$ and K$^{+}$ cations can be treated as
small non-polarizable hard ions and they are repelled from the
aqueous interfaces. These results suggest that the electrolyte
aqueous solution surfaces are more complex than the currently
prevalent theoretical and experimental understandings.
\end{abstract}

\maketitle

\section{Introduction}

Recently, aqueous solution surfaces of simple electrolytes have
attracted much attention for its importance in the broad chemical,
environmental and biological processes at the molecular interfaces
and
membranes.\cite{AllenChemRev,GarrettScience,TobiasScience08,TobiasChemRev,JungwirthARPC,JungwirthFaraday2009}
Conventionally, it has been widely believed that the aqueous
solution surfaces were empty of ions. This classical picture was
based on the macroscopic surface tension measurement of the
electrolyte solutions, and was theoretically described by the
negative excess of the ions from the thermodynamic Gibbs adsorption
equation, as well as the microscopic image charge repulsion
theory.\cite{AdamsonBook,OnsagerJCP528} Recently, new findings in
the atmospheric chemical research, direct surface measurement
techniques and molecular dynamics (MD) simulations have challenged
this classical picture of ions depletion at the electrolyte solution
surface region since Gibbs and
Onsager.\cite{HuJPC8768,GhosalScience,TobiasScience,AllenChemRev,TobiasChemRev,SaykallyARPS,BianPCCP}

Consequently, theoretical simulations and experimental techniques
have been employed to provide answers to the two equally important
questions about the electrolyte aqueous solution surface. One is how
the anions and cations are spatially distributed in the surface
region, and the other is how the interfacial water molecules are
affected by this ion distribution. To answer the former question
requires techniques which can directly measure the presence and
profile the depth of the ion distribution at the liquid surface;
while to answer the later requires techniques which can directly
measure the presence and change of the interfacial water molecules.
The convergence of and the correlation between the two answers may
provide a through picture for the understanding of the electrolyte
aqueous solution surface. The fact is that the conclusions of most
of the recent theoretical and experimental studies converged on the
picture that the anions, especially the larger and more polarizable
anions, are presented at the interfacial region of various simple
inorganic salt solutions.

Molecular dynamics (MD) simulation results using the polarizable
force fields concluded for significant enrichment of the anions,
such as the larger and more polarizable I$^{-}$ and Br$^{-}$, at the
interfacial
region.\cite{TobiasChemRev,TobiasFeatureArticle6361,Tobias7617,DangChemRev,Tobias10468}
MD simulation also revealed that the anions and cations should be
segregated for the electrolytes consist of more polarizable anions
at the electrolyte aqueous solution surface, with the double layer
structure in which the anions enriched at the top surface layer and
the cations resided below.\cite{JungwirthARPC} Except for some
voices of discontent,\cite{Richmond5051,SloutskinJCP054704} the
majority of the experimental studies on the electrolyte aqueous
solution surfaces, including the studies with X-ray photoelectron
spectroscopy (XPS),\cite{GhosalScience,WinterChemRev2006} X-ray
fluorescence\cite{PadmanabhanPRL086105} and nonlinear optical
spectroscopy techniques, such as the resonant second harmonic
generation (resonant
SHG)\cite{SaykallyCPL51,SaykallyCPL46,Saykally10915,
Saykally7976,SaykallyJACS,SaykallyARPS} and sum-frequency generation
vibrational spectroscopy,
(SFG-VS)\cite{AllenSodiumHalide2252,ShultzJPCB585,Motschmann4484,Motschmann2099,AllenJPCB8861,ShenJACS13033}
lent their support to this new physical picture with the more
polarizable anions enriched at the electrolyte aqueous solution
surface.

Since in this new picture the more polarizable anions are expected
to be present at the electrolyte aqueous solution surface, while the
small non-polarizable hard cations, such as  Na$^{+}$ and K$^{+}$
etc., and anions, such as F$^{-}$, are expected to be repelled from
the aqueous solution surface, experimental studies have been focused
mainly on either to prove the above
conclusions,\cite{GhosalScience,SaykallyCPL51,SaykallyCPL46,Saykally10915,
Saykally7976,SaykallyJACS,SaykallyARPS,Motschmann4484,Motschmann2099,HemmingerPCCP4778}
or to investigate the specific anion effects on the interfacial
water
molecules.\cite{AllenSodiumHalide2252,Richmond5051,AllenChemRev,ShenYRPRL096102,BianPCCP}
Therefore, to our best knowledge, so far there has been no
systematic investigation on the cation effects on the interfacial
water molecules at the electrolyte aqueous solution
surfaces.\cite{TobiasChemRev,JungwirthARPC,JungwirthFaraday2009}

Recently, we developed and employed the surface sensitive
non-resonant SHG technique to measure the polarization dependent SHG
response from the interfacial water molecules at the NaF, NaCl, and
NaBr salt solution surfaces.\cite{BianPCCP,HongfeiPCCP} In this
study, in order to ensure measurement of small changes in the SHG
signal, we first developed effective procedures to monitor and
remove the impurities in the salt solution samples. Then, the
quantitative polarization analysis of the measured SHG data showed
that the average orientation of the interfacial water molecules
changed slightly with the increase of the bulk concentration of the
NaF, NaCl and NaBr salts from that of the neat air/water interface.
The observed significant SHG signal increase with the bulk salt
concentration was attributed to the overall increase of the
thickness of the interfacial water molecular layer, following the
order of NaBr $>$ NaCl $\sim$ NaF. The absence of the
electric-field-induced SHG (EFISHG) effect indicated that the
electric double layer at the salt aqueous solution surface is much
weaker than that predicted from the molecular dynamics (MD)
simulations. These results provided quantitative data to the
specific anion effects on the interfacial water molecules of the
electrolyte aqueous solution, not only for the larger and more
polarizable Br$^{-}$ anion, but also for the smaller and less
polarizable F$^{-}$ and Cl$^{-}$ anions.

The most intriguing thing in this study was that we were able to
quantitatively measure weak concentration dependent F$^{-}$ anion
effect of the NaF salt on the interfacial water
molecules,\cite{BianPCCP} even though this F$^{-}$ anion effect was
expected to be insignificant from the recent MD
simulations.\cite{TobiasChemRev,HemmingerPCCP4778} This immediately
led us to look at whether the F$^{-}$ anion effect of the KF salt
aqueous solution surface can also be observed. Moreover, there is
also one known advantage to investigate the KF aqueous solution
because this can extend the bulk concentration range of the F$^{-}$
anion. The saturation solubility for NaF in water at 20$^{\circ}$C
is only 0.98M (4.13g NaF per 100g water), while the saturation
concentration in KF is 15.4M (89.8g KF per 100g
water).\cite{CRCHandBook} Surprisingly, we not only observed the
significant F$^{-}$ anion effect of the KF salt aqueous solution
surface up to much higher bulk concentration as expected, we also
found that the orientational parameter of the interfacial water
molecules went to the opposite direction to that of the NaF aqueous
solution surface as the bulk concentration of NaF and KF salt
increase. Therefore, we started doing experiment also on the surface
of the KCl and KBr salt aqueous solutions, and we observed
distinctive specific Na$^{+}$ and K$^{+}$ cation effects on the
interfacial water molecules at the NaF, NaCl, NaBr, KF, KCl, and KBr
salt solution surfaces.

In the following sections, we shall report the non-resonant SHG
measurement and analysis of the specific Na$^{+}$ and K$^{+}$ cation
effects on the interfacial water molecules at these six solution
surfaces. We shall show that while the thickness of the interfacial
water layer at the surfaces of all the six salt solutions increases
as the bulk concentration increase, the orientational parameter D
values of the interfacial water molecules changed to opposite
directions for the aqueous solutions of the three sodium salts
versus the aqueous solutions of the three potassium salts. Moreover,
the abilities to increase the thickness of the interfacial water
layer were found to be in the following order: KBr $>$ NaBr $>$ KCl
$>$ NaCl $\sim$ NaF $>$ KF. These specific Na$^{+}$ and K$^{+}$
cation effects, as well as the F$^{-}$, Cl$^{-}$ and Br$^{-}$ anion
effects, indicate that the electrolyte aqueous solution surfaces are
likely to be more complex than what the currently prevalent
theoretical and experimental understandings can offer.

\section{Theoretical background for the SHG data treatment}

The theoretical tools used in this report shall fully follow the
detailed description in the previous published paper, since this
work is an extension from the previous study.\cite{BianPCCP} Here we
only present what is necessary to understand the analysis of the SHG
data in this report.

The SHG intensity $I(2\omega)$ reflected from the electrolyte
aqueous interface is given
by\cite{RaoYiJCP,ShenZhuangPRB,ZhengDSIRPC2008}

\begin{equation}
I(2\omega ) = \frac{{32\pi ^3 \omega ^2 \sec ^2 \beta }} {{c_0^3 n_1
(\omega )n_1 (\omega )n_1 (2\omega )}}\left| {\chi _{eff} }
\right|^2 I^{2}(\omega )\label{intensity}
\end{equation}

\noindent Here $I(\omega)$ is the incoming laser intensity, $c_0$ is
the speed of the light in the vacuum, $\beta$ is the incident angle
from the surface normal, defined as $z$ axis in the laboratory
coordinates system $(x,y,z)$, $n_{1}(\omega_{i})$ is the refractive
index at the frequency $\omega_{i}$ of the medium in which the laser
beam propagates. $\chi_{eff}$ is the effective macroscopic second
order susceptibility. In the SHG experiment, because there are only
one incident laser beam and one out-going signal beam, there are
three independent polarization combinations for the $\chi_{eff}$:
namely the s-in/p-out ($\chi _{eff,sp}$), the $45^{\circ}$-in/s-out
($\chi _{eff,45^{\circ}s}$) and the p-in/p-out ($\chi _{eff,pp}$).
Here $p$ denotes the polarization in the incident plane, while $s$
the polarization perpendicular to the incident plane. With the
microscopic local-field factors incorporated implicitly into the
tensorial Fresnel factor $L_{ii}$s, the $\chi_{eff}$s
are\cite{ShenZhuangPRB,ZhengDSIRPC2008}

\begin{eqnarray}
\chi_{eff,sp}^{(2)}&=&L_{zz}(2\omega)L^2_{yy}(\omega)sin\beta\chi_{zyy}\nonumber
\\
\chi_{eff,45^{\circ}s}^{(2)}&=&L_{yy}(2\omega)L_{zz}(\omega)L_{yy}(\omega)sin\beta\chi_{yzy}\nonumber
\\
\chi_{eff,pp}^{(2)}&=&+L_{zz}(2\omega)L^2_{xx}(\omega)sin\beta{cos^2\beta}\chi_{zxx}\nonumber\\
& &
-2L_{xx}(2\omega)L_{zz}(\omega)L_{xx}(\omega)sin\beta{cos^2\beta}\chi_{xzx}\nonumber\\
& &
+L_{zz}(2\omega)L^2_{zz}(\omega)sin^3\beta\chi_{zzz}\label{xeff-chi}
\end{eqnarray}

\noindent in which the $\chi _{ijk}$s are the seven non-zero
susceptibility tensors, i.e. $\chi_{zzz}$, $\chi_{zxx}=\chi_{zyy}$,
$\chi_{xzx}=\chi_{xxz}=\chi_{yzy}=\chi_{yyz}$, of the rotationally
symmetric interface along the interface normal $z$, with $x$ in the
incident plane.

In the SHG measurement, the polarization is usually fixed at either
$p$ or $s$, while the polarization angle $\Theta$ of the incident
laser beam is varied for the full $360^{\circ}$ using a half
waveplate. Then, one has\cite{ZhangWKJCP224713,HongfeiPCCP}

\begin{eqnarray}
I_p  &\propto& |\chi_{eff, p}|^{2}=|\chi _{eff,pp}\cos ^2 \Theta  + \chi _{eff,sp}\sin ^2 \Theta |^2 \nonumber \\
I_s  &\propto&  |\chi_{eff, s}|^{2}=|\chi _{eff,45^{\circ}s}\sin
2\Theta |^2 \label{p-s}
\end{eqnarray}

In the case of non-resonant SHG, each of the $\chi _{eff,pp}$, $\chi
_{eff,sp}$ and $\chi _{eff,45^{\circ}s}$ terms is real, and their
relative values can be obtained from fitting the experimental data
with the Eq.\ref{p-s} by knowing the relative signs between them.
The relative signs can be determined with the measurement of the SH
signal in the $45^{\circ}$ polarization as described
elsewhere.\cite{HongfeiPCCP} It was also shown that using the
Eqs.\ref{p-s}, $\chi _{eff,pp}$, $\chi _{eff,sp}$ and $\chi
_{eff,45^{\circ}s}$ terms can be most accurately determined from the
polarization dependent SHG data.\cite{HongfeiPCCP}

In the previous study, we showed that the contributions from the
ions and from the electric field induced SHG were negligible for the
non resonant SHG signal from the electrolyte aqueous solution
surface. Therefore, only the contribution from the interfacial water
molecules need to be considered. For molecules with the $C_{2v}$
symmetry, such as the water molecule, there are seven nonzero
microscopic polarizability tensor elements
$\beta_{ijk}(i,j,k=a,b,c)$ in molecular coordinates system
$(a,b,c)$:  namely, $\beta_{ccc}, \beta_{caa}, \beta_{cbb},
\beta_{aca}, \beta_{aac}, \beta_{bcb}$ and $\beta_{bbc}$, with
$\beta_{aca}=\beta_{aac}$ and $\beta_{bcb}=\beta_{bbc}$ for the SHG
process. When the optical frequency is far below the resonance,
$\beta_{ccc}$ is usually negligible, by defining $R=(\beta_{caa}+
\beta_{cbb})/(\beta_{aca}+ \beta_{bcb}$), one
has,\cite{ZhangWKJCP224713,HongfeiPCCP}

\begin{eqnarray}
\frac{{\chi _{zxx} }} {{\chi _{xzx} }} &=& \frac{{(R - 2)D + (R +
2)}}
{{ - RD + (R + 2)}} \nonumber \\
\frac{{\chi _{zzz} }} {{\chi _{xzx} }} &=& \frac{{2(R + 2)D - 2(R +
2)}} {{ - RD + (R + 2)}} \label{ratio}
\end{eqnarray}

Here $D=\langle cos\theta\rangle/\langle cos^3\theta\rangle$ is the
orientational parameter with $\theta$ as the tilt angle of the water
dipole axis from the interface normal, $\langle \rangle$ denotes the
ensemble average over the whole orientational distribution. Because
the two ratios $\chi _{zxx}/\chi _{xzx}$ and $\chi_{zzz}/\chi
_{xzx}$ can be accurately obtained from the polarization SHG
measurement, the $R$ and $D$ values can be readily obtained with
good accuracy, using the
Eqs.\ref{ratio}.\cite{ZhangWKJCP224713,HongfeiPCCP}

Now with the orientational parameter $D$ known, the relative number
of water molecules contributing to the measured SHG signal from the
interface of the different electrolytes as well as different
electrolyte concentrations can be calculated from the changes of the
corresponding SHG intensities. Simply to put it,

\begin{eqnarray}
I_{2\omega}  \propto |\chi _{eff} |^2  \propto |N_s\chi _{eff,0}|^2
= N_s^{2}|r(\theta)|^2\label{numberdensity}
\end{eqnarray}

Here $N_{s}$ is the interfacial number density of the water
molecule, $\chi _{eff,0}$ is the average effective susceptibility
per molecule. Here $\chi _{eff,0}$ can also be written into the per
molecule orientational functional $r(\theta)$, which is directly
related to the orientational parameter D.\cite{RaoYiJCP,HongfeiPCCP}

\section{Experimental details}

The procedures to pretreat the salt sample and purify the salt
solution were described in detail previously and should be referred
when in doubt.\cite{BianPCCP} The necessary experimental detail and
the information for sample preparation are described below.

\subsection{SHG measurement}

The SHG experimental setup is as in the previous
reports.\cite{ZhangWKJCP224713,RaoYiJCP,BianPCCP} A broadband
tunable mode-locked femtosecond Ti:Sapphire laser (Tsunami 3960C,
Spectra-Physics) is used for the reflected-geometry SHG measurement.
Its high-repetition rate (82 MHz) and short pulse width (80 fs) make
it suitable for detection of the weak second-harmonic signals. Its
long term power and pulse-width stability also make it easy for
quantitative analysis of the SHG data.\cite{RaoYiJPCA7987} The 800
nm fundamental laser beam is focused on the solution interface at
the incident angle of $\beta$=70$^{\circ}$, and the SHG signal at
400 nm is detected with a high-gain photomultiplier tube for single
photon counting (R585, Hamamatsu) and a photon counter (SR400,
Stanford Research Systems). Typically the dark noise level is less
than 1 counts/s, better than reported
previously.\cite{ZhangWKJCP224713} The typical laser power is 500
mW. The efficiency of the detection system for the p polarization is
1.23 times of that for the s polarization. A cylindrical Teflon
beaker ($\phi$50x6mm) is used to host the solution. The optical
polarization control and the SHG data acquisition are programmed and
collected with a PC. It takes about 150s to 300s to collect the
polarization curve. The room temperature is controlled at
22.0$\pm$1.0$^{\circ}$C, and the humidity in the room is controlled
around 40$\%$.

\subsection{Sample preparation}

In the experiments, the liquid water was purified with a Millipore
Simplicity 185 ($18.2{M}{\Omega}{\cdot}cm$) from double distilled
water. The KF, KCl and KBr used were purchased from the
Sigma-Aldrich (ACS reagent grade, $99\% +$ purity). Before the
preparation of the salt solution, the salts were baked at around
$500^{\circ}C$ for more than 6 hours in order to remove the organic
impurities. The glassware was cleaned with hot chromic acid and then
rinsed with Millipore water for several times. The solutions were
filtered with a syringe with a 0.22$\mu$m membrane filter made of
PVDF material by Millipore (Durapore GVWP01300) to remove the
possible insoluble particles existing in the solution. The SFG-VS
spectral and SHG fluctuation measurement are used to monitor the
contaminates.

\section{Results and Discussion}


\subsection{Quantitative analysis of the polarization dependent non-resonant SHG data}

In the previous study,\cite{BianPCCP} we showed that the
non-resonant SHG signal from the aqueous solution surface of simple
electrolytes was originated from the surface water molecules, and
the contribution from the surface ions was negligible. We also
showed that the contribution to the non-resonant SHG signal through
the electric field induced second harmonic generation (EFISHG)
mechanism possibly from the surface potential due to the surface
charge separation was negligible. Careful sample treatment and
purification procedures were also developed to ensure data
reproducibility. Here we just followed the experimental and data
analysis procedures established in the previous literature, from
which the accuracy and reproducibility of the non-resonant SHG data
and the quantitative analysis of the SHG data were
assured.\cite{BianPCCP,HongfeiPCCP}

\begin{figure}[h]
\begin{center}
\includegraphics[height=9cm]{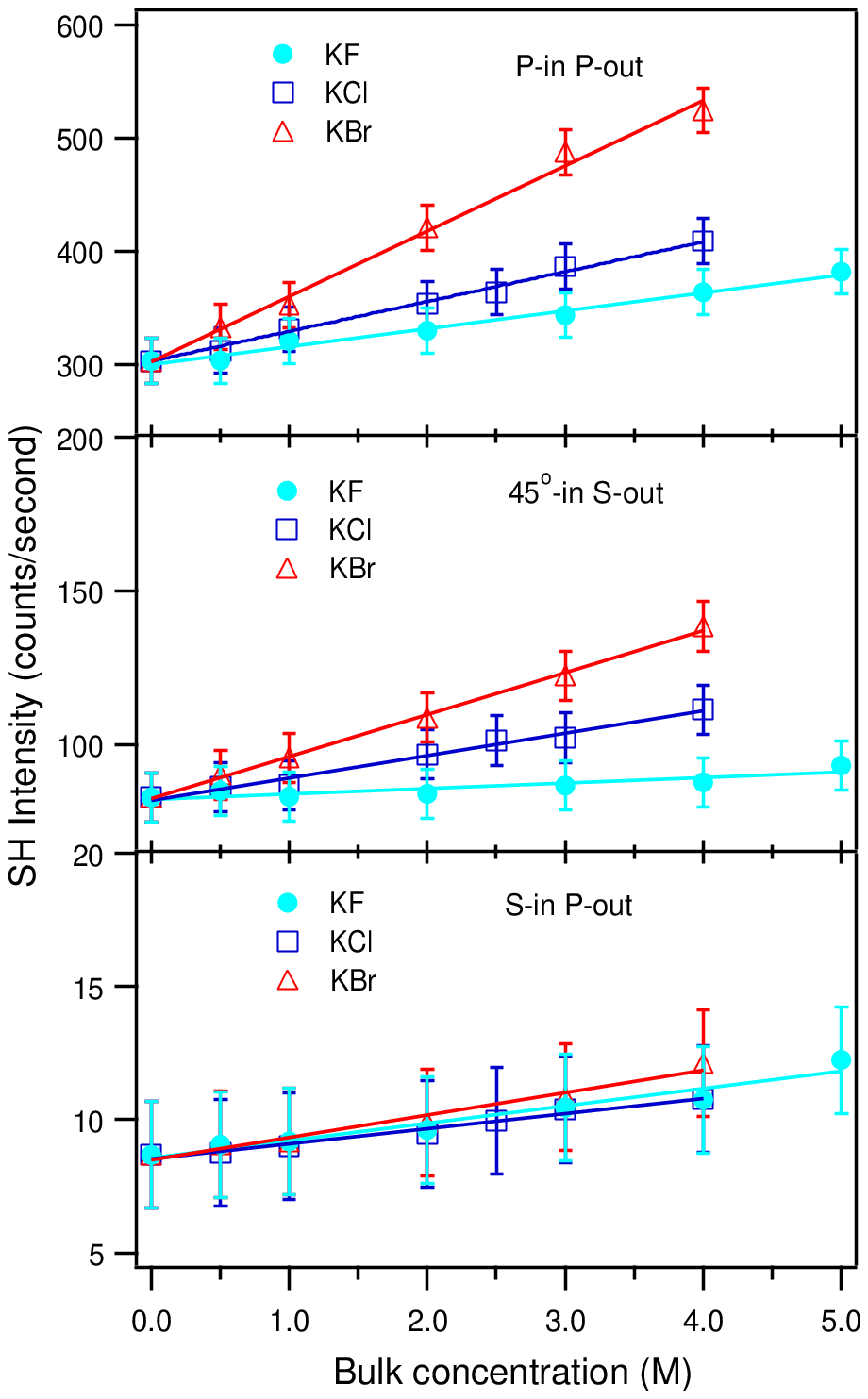}
\hspace{0.3cm}
\includegraphics[height=9cm]{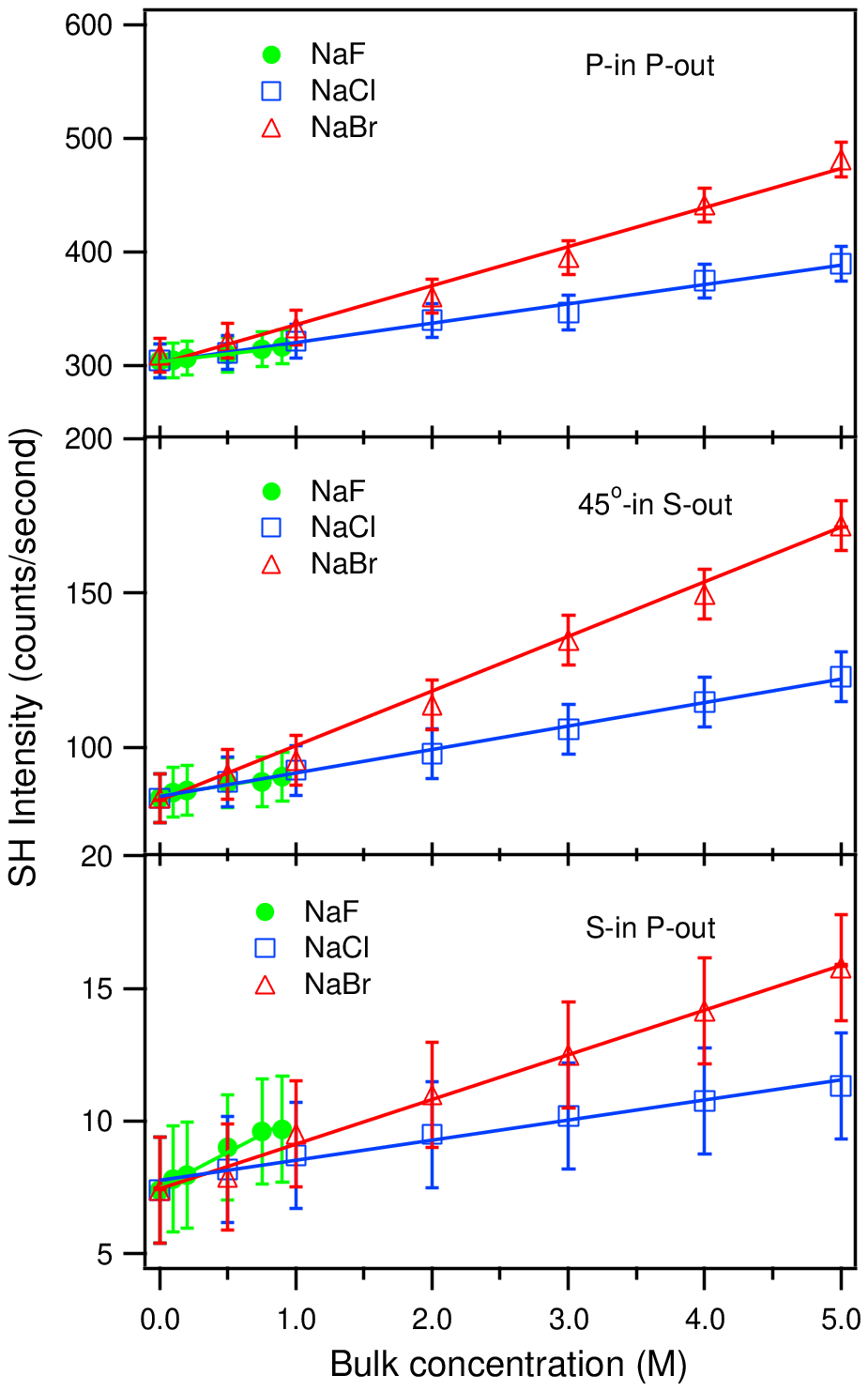}
\caption{Polarization and concentration dependent SHG (400nm) data:
a. left panel, KF, KCl and KBr solution surfaces; b. right panel,
NaF, NaCl and NaBr solution surfaces. These data were obtained by
fitting the s and p detection data measured for each concentration
and each salt solution. The procedure used was fully described
previously,\cite{ZhangWKJCP224713,BianPCCP} and it can ensure the
accuracy of the data for each polarization combination. The
45$^{\circ}$-in/s-out data need to be scaled up by the factor of
1.23 because of the different detection efficiency for the s and p
polarized SHG signal at 400nm in the detection setup in this
experiment. The data for NaF, NaCl and NaBr are identical to the
data in our previous article,\cite{BianPCCP} and the scale is
changed to be the same as the KF, KCl and KBr data for better visual
comparison.}\label{KSHIntensity}
\end{center}
\end{figure}

Fig. \ref{KSHIntensity} shows the non-resonant SHG intensities for
the NaF, KF, NaCl, KCl, NaBr and KBr solution surfaces in the three
independent polarization combinations, i.e. p-p, 45$^{\circ}$-s and
s-p polarization combinations, at different concentrations. The data
for the NaF, NaCl and NaBr were published in our previous
work.\cite{BianPCCP} The saturation concentrations of KF, KCl and
KBr at the 20$^{\circ}$C are 15.4M (89.8g KF per 100g water), 4.6M
(34.2 g KCl per 100g water), and 6.3M (65.3g KBr per 100g water),
respectively.\cite{CRCHandBook} In comparison, the saturation
concentration of the NaF, NaCl and NaBr at 20$^{\circ}$C are
0.98M(4.13 g NaF per 100 g water), 6.2 M (36.0 g NaCl per 100 g
water), and 9.2 M (94.6 g NaBr per 100 g water).\cite{CRCHandBook}
As mentioned above, the large saturation concentration for KF
enabled us to go to higher bulk concentration with the F$^{-}$ anion
by using the KF salt; while with the NaF, the saturation
concentration is only 0.98M. In the experiment, the highest bulk
concentration for KF, KCl, and KBr are 5M, 4M and 4M, respectively.

It is clear from the Figure \ref{KSHIntensity} that for all the
salts, the SHG signal increased almost linearly with the increase of
the bulk electrolyte concentration, and the general behaviors for
the potassium salts are similar to those of the sodium salts, with
the biggest slope for the KBr and smallest slope for the KF.
Therefore, this suggests that similar to the NaF, NaCl and NaBr
salts,\cite{BianPCCP,AllenSodiumHalide2252} KF, KCl and KBr salts
also resulted in the increase of the thickness of the interfacial
water molecular layer. However, a closer examination of the SHG data
revealed apparent differences between the NaF/KF, NaCl/KCl and
NaBr/KBr pairs, indicating non-negligible specific Na$^{+}$ and
K$^{+}$ cation effects on the non-resonant SHG signal of the
interfacial water molecules. For example, in the Figure
\ref{KSHIntensity}, the slopes of the increase of the p-p signal for
the KF, KCl and KBr are larger than the slope for the NaF, NaCl and
NaBr, respectively; while for the 45$^{\circ}$-s and s-p data showed
completely the opposite trend. This relative differences between the
intensities of different polarization combinations usually indicate
possible differences of orientational order of the surface molecules
contributing to the SHG
signal.\cite{RaoYiJCP,HongfeiPCCP,ShenZhuangPRB}

\begin{figure}[h]
\begin{center}
\includegraphics[width=7cm]{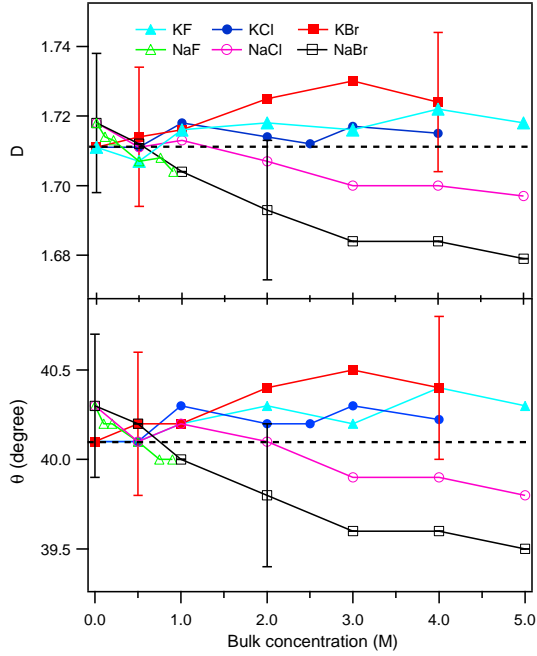}
\caption{The orientational parameters D (top) and the apparent
orientational angles $\theta$ (bottom) for the KF (solid triangle),
KCl(solid circle), KBr (solid square), NaF (open triangle),
NaCl(open circle) and NaBr (open square) solution interfaces. The
data for KF, KCl and KBr are from Table \ref{WaterFittingResults}.
The data for NaF, NaCl and NaBr were from the Table 1 in the
previously published paper.\cite{BianPCCP} It is clear that the D
values for the neat water and low bulk concentration for the sodium
halides and the potassium halides data series were well within the
experimental error; while at higher concentrations the D values for
the two series went to different directions.}\label{DThetaValueNaK}
\end{center}
\end{figure}

The differences of the orientational order parameter D values are
clearly presented in the Figure \ref{DThetaValueNaK}. The D values
were obtained by using Eqs. \ref{xeff-chi} and Eqs. \ref{ratio}, and
the calculated results for the KF, KCl and KBr solutions are listed
in the table \ref{WaterFittingResults}. In the table
\ref{WaterFittingResults}, the calculated molecular polarizability
ratio $R=(\beta_{caa}+ \beta_{cbb})/(\beta_{aca}+ \beta_{bcb}$)
values for all three salts at different bulk concentrations were all
the same within the small error bar as the value calculated from the
SHG data for the neat air/water interface.\cite{ZhangWKJCP224713}
These values are also the same as those obtained for the NaF, NaCl
and NaBr aqueous solution surfaces as reported
previously.\cite{BianPCCP} This further confirmed that the
assumption that the non-resonant SHG observed from the KF, KCl and
KBr aqueous solution surfaces were indeed from the interfacial water
molecules. The D and $\theta$ values for the NaF, NaCl and NaBr
aqueous solution surfaces, which were published
previously,\cite{BianPCCP} are also plotted in the Figure
\ref{DThetaValueNaK} for comparison. Even though the D values only
changed slightly at different bulk salt concentration, the
difference is clearly bigger than the error bar as the bulk salt
concentration increase. It is interesting to see that the D values
for the potassium salt series are apparently larger than the D
values for the sodium salt series. This also confirms the
qualitative surmise above based on the different trends for the
polarization dependence of the non-resonant SHG signal for the KX
(X=F, Cl or Br) salt series and the NaX salt series.

In the Figure \ref{DThetaValueNaK}, the $\theta$ value is the
apparent average tilt angle between the dipole vector of water
molecule and the surface normal calculated using the assumption of a
$\delta$ distribution function for this tilt angle. The assumption
for a $\delta$ distribution function of $\theta$ is certainly not
correct. However, as discussed intensively in the literature,
because within a certain range of the width for the orientational
distribution, this $\theta$ value can be used to represent the
center value for the orientational distribution and can be used to
simulate the orientational dependent total SHG susceptibility, i.e.
the orientational functional
$r(\theta)$.\cite{RaoYiJCP,GanWeiJPCCEthanol,WangIRPC} This point
shall be further illustrated below.

With the orientational parameter known, the relative surface density
$N_{s}$ of the interfacial water molecules which contribute to the
measured non-resonant SHG signal can be readily calculated using the
Eqs. \ref{numberdensity} These values for the KX salts are also
listed in the Table \ref{WaterFittingResults}. Figure
\ref{WaterNumberNaK} shows the $N_{s}$ values for the NaF, KF, NaCl,
KCl, NaBr, and KBr salt aqueous solution surfaces. The $N_{s}$
values for the three NaX salts are the same as from the previously
published results.\cite{BianPCCP} It is clear the $N_{s}$ for each
salt increases almost linearly with the bulk salt concentration. As
discussed previously,\cite{BianPCCP} since the probing area of the
laser is the same, the increase of the $N_{s}$ represents the
increase of the average depth of the water molecules in the
interfacial region which contributes to the non-resonant SHG signal.
Now we can fit the $N_{s}$ versus the bulk concentration of each of
the salts data with a linear function of $N_{s}= N_{s,0} + kC$. Here
$N_{s,0}$ = 1 denotes the surface density of the neat air/water
interface, $C$ is the bulk salt concentration, and the slope $k$
represents the ability to increase the thickness of the surface
water molecules. The fitting results of $k$ values are listed in the
Table \ref{NumberPowerFitting}. The unit for $k$ is M$^{-1}$.

According to the $k$ values in the Table \ref{NumberPowerFitting},
the abilities to increase the surface water thickness for the six
salts follow the following order: KBr $>$ NaBr $>$ KCl $>$ NaCl
$\sim$ NaF $>$ KF. Clearly, both the anions and cations contributes
to the increase of $N_{s}$. In the previous study with the NaF, NaCl
and NaBr salts, the order was NaBr $>$ NaCl $\sim$ NaF, and mostly
the effect was attributed to the ability to be polarized for the
larger anions. However, here we clearly have KBr $>$ KCl $>$ KF, and
one can also conclude that even though there are differences on the
abilities to increase the aqueous surface thickness for the
individual ions, the abilities for the cations and anions are
generally with comparable order of magnitudes. Therefore, the cation
effects can no longer be ignored. These findings on the specific
anion and cation effects from the quantitative analysis of the
non-resonant SHG data imply that the electrolyte aqueous solution
surface can be more complex than previously thought.

\subsection{The structure and reorientation of the interfacial water molecules at the
electrolyte solution surfaces}

There have been many theoretical and experimental studies on the
structure of the interfacial water molecules. The vibrational
spectral feature around 3700cm$^{-1}$ of the free O-H stretching
vibration was first observed by Shen and co-workers in 1993 using
the surface specific sum frequency generation vibrational
spectroscopy (SFG-VS).\cite{ShenYRPRL2313} Subsequent SFG-VS
experimental studies and quantitative analysis further concluded
that this water molecule with the free O-H straddles across the
air/water interface with its dipole vector almost parallel to the
interface, i.e. on average the free O-H tilted from the surface
normal around
30-35$^{\circ}$.\cite{RichmondChemRev2002,ShenChemRev2006,GanWeiJCP114705,GanWeiCJCP20,TyrodeJPCB329}
This latter conclusion was also supported by molecular dynamics (MD)
and Monte Carlo (MC)
simulations.\cite{Wilson4873,YangF109,Benjamin2083,Besseling11610,
Taylor11720,Sokhan625,Fradin871,Morita371,Perry8411,Kuo658,Jaqaman926}
SFG-VS experiment also revealed that the SFG spectra in the
3200-3600cm$^{-1}$ region were from the contributions of the
hydrogen bonded water molecules slightly below the topmost water
layer at the air/water interface, and these water molecules had both
of their hydrogen atoms hydrogen
bonded.\cite{ShenYRPRL2313,GanWeiJCP114705} These are the
non-straddle-type water molecules at the air/water interface, and
the SHG signal from the air/water interface comes mainly from the
non-straddle-type water molecules.\cite{ZhangWKJCP224713}

The SFG-VS studies by Allen and Richmond and their co-workers on the
NaF, NaCl, NaBr and NaI aqueous solution surfaces concluded that the
free O-H spectra was almost undisturbed with the addition of the
salts, while the hydrogen bonded water spectra were variously
affected.\cite{AllenSodiumHalide2252,Richmond5051} Using the more
informative phase-sensitive sum-frequency spectroscopy technique,
Shen \textit{et al.} recently investigated the hydrated I$^{-}$
anion effects on the interfacial water
structure.\cite{ShenYRPRL096102} Their conclusion was that the
presence of I$^{-}$ anion near the interface region may not disturb
the water molecular structure at the topmost surface layer, while it
can reorient the water molecular in the subphase. These conclusions
are consistent with our non-resonant SHG results on the interfacial
water molecules at these electrolyte aqueous solution surfaces.
Moreover, our SHG results provides not only quantitative measurement
on the slight reorientation of the interfacial water molecules, but
also quantitative measurement on the increase of the average
thickness of the interfacial water layer, for the electrolyte
aqueous solution surfaces.

Here we would like to further discuss the different reorientation
effects on the interfacial water molecules by the different anions
and cations.

Firstly, the change of the values of the orientational parameter D
in the whole bulk concentration range is small. If we do not compare
the results for the KX series with the NaX series, we may conclude
that there is essentially no apparent change of the D values for the
KX series. This indicates that the reorientation effect for the KX
series is rather insignificant. Therefore, the significant increase
of the SHG signal is essentially not the result of water molecule
reorientation.

Secondly, from the Figure \ref{DThetaValueNaK}, the direction of the
reorientation effect is mainly determined by the Na$^{+}$ and
K$^{+}$ cations, instead of the anions. This finding is intriguing
because neither the classical picture nor the recently revised
picture can be used to predict this significant cation effect. In
the classical theory, both the anion and cations are thought to be
repelled from the aqueous solution surface. The recent MD
simulations which predicted enrichment of more polarizable anions
also negated the presence of the non-polarizable cations, as well as
the non-polarizable anions, at the aqueous solution
surface.\cite{TobiasChemRev} The results here suggested that
probably even for the much less polarizable cations and anions can
still be present or enriched at the interface region. It is probable
that in order to keep the neutrality of the anion enriched interface
region, more cations are attracted to the surface region. This
picture is certainly consistent with the fact we reported previously
that there was insignificant EFISHG contribution to the total SHG
signal.\cite{BianPCCP} However, in order to find reconciliation with
the current picture of charge segregation from the MD simulation
results, much effort are still needed.

With the known R values, the orientation dependent per molecule
$\chi_{eff,0}$ values in different polarization combinations can be
simulated using the Eqs.\ref{xeff-chi} and Eqs.\ref{ratio}. When the
$N_{s}$ value was known, the $\chi_{eff,0}$ value relative to the
neat air/water interface value $\chi_{eff,water}$ can be calculated
with the Eqs.\ref{numberdensity}. Comparison of these two values can
help us understand how the SHG signal changes with the averaged
molecular orientation of the water molecules at the electrolyte
aqueous solution surfaces. Figure \ref{XeffSimulation} shows the
simulation results of the $\chi_{eff,0}$ values in the p-p,
45$^{\circ}$-s and s-p polarization combinations using R=0.69 and
$(\beta_{aca}+ \beta_{bcb})$=1. From the curves in the Figure
\ref{XeffSimulation}, we can see that if the average orientation
angle $\theta$ becomes slightly larger than the neat air/water
interface value $\theta$=40.1$^{\circ}$, the $\chi_{pp,0}$ value
shall become larger, while the $\chi_{45^{\circ},0}$ and
$\chi_{sp,0}$ values shall become smaller. This is indeed the case
for the KF, KCl and KBr aqueous solution surfaces from the SHG
measurement. On the other hand, if the average orientation angle
$\theta$ becomes slightly smaller than the neat air/water interface
value $\theta$=40.1$^{\circ}$, the $\chi_{pp,0}$ value shall become
smaller, while the $\chi_{45^{\circ},0}$ and $\chi_{sp,0}$ values
shall become larger. This latter case is consistent with the SHG
data for the NaF, NaCl and NaBr aqueous solution surfaces.

In the Figure \ref{XeffSimulation}, we also presented the simulation
of the $\chi_{eff,0}$ values when the orientational distribution
function is not the $\delta$ function. It is clear that when the
full width at the half maximum (FWHM) of the orientational
distribution function is 30$^{\circ}$, the trend of $\theta_{0}$
dependent change of each $\chi_{eff,0}$ component remained the same
in the vicinity of $\theta$=40.1$^{\circ}$. For example, just like
the $\delta$ distribution case, when the average orientation angle
$\theta$ becomes slightly larger than the neat air/water interface
value $\theta$=40.1$^{\circ}$, the $\chi_{pp,0}$ value shall become
larger, while the $\chi_{45^{\circ},0}$ and $\chi_{sp,0}$ values
shall become smaller. Our simulation further demonstrated that as
long as FWHM does not exceed 65$^{\circ}$, this trend shall remain
the same. Therefore, the conclusions that the NaX salt series
slightly reorient the interfacial water molecules to smaller average
tilt angle and the KX salt series slightly reorient the interfacial
water molecules to larger average tilt angle remained unaffected
when the orientation of the interfacial water molecules possess a
broad range of distribution width. This is another example to
illustrate that in the polarization and orientation analysis of SHG,
as well as SFG, the orientational order can be understood even
though the orientational distribution is in a broad
range.\cite{GanWeiJPCCEthanol} Thus, the small shift of the average
orientation of the interfacial molecules, as representatively shown
in the Figure \ref{SurfaceWaterStructure}, can be accurately
determined through polarization and orientational analysis of the
SHG/SFG data.

The unexpected specific cation effect on the reorientation of the
interfacial water molecules by the Na$^{+}$ and K$^{+}$ cations as
determined here indicates complex interactions between the ions and
the water molecules at the electrolyte aqueous solution surfaces. It
is rather surprising to see that even though the F$^{-}$, Cl$^{-}$
and Br$^{-}$ anions influenced the extent of the reorientation
effect, they do not control the direction of the reorientation.
These are surprising results from our current understandings on the
ion adsorption to the aqueous surfaces.

Based on these different cation effects on the hydrogen bonded water
molecules in the surface region, we surmised that there might be
some observable effects on the hydrogen bonding O-H vibrational
spectra for these interfacial water molecules. This prompted us to
measure the SFG vibrational spectra of the interfacial water
molecules in the 3000-3800cm$^{-1}$ region at the NaF and KF aqueous
solution surfaces. The SFG-VS spectra of the NaF aqueous solution
surfaces have been investigated by various research groups. However,
these data were not quite consistent with each
other.\cite{AllenSodiumHalide2252,AllenChemRev,Richmond5051,BianPCCP}
Moreover, to our knowledge, there has been no SFG-VS measurement in
the literature on the water molecules at the KF aqueous solution
surface, partly because the cation effects between the Na$^{+}$ and
K$^{+}$ on the surface water molecules were expected to be
negligible in the best of the current
understandings.\cite{TobiasChemRev,HemmingerPCCP4778,TobiasFeatureArticle6361}

We measured and quantitatively analyzed the SFG-VS spectra of the
neat air/water interface and other water surfaces in the past few
years.\cite{GanWeiCJCP20,GanWeiJCP114705,ZhangZhenPCCP} Based on
these successful experiences, we measured the SFG-VS spectra of the
NaF and KF aqueous solution surfaces at different bulk
concentrations. We found that the NaF water SFG-VS spectra were
consistent with the reported spectra by the Richmond
group,\cite{Richmond5051,BianPCCP} and we also observed distinctive
differences in the hydrogen bonding O-H stretching region in
3000-3600cm$^{-1}$ for the NaF and KF aqueous solution surfaces.
These results firmly confirmed the specific cation effects between
the Na$^{+}$ and K$^{+}$ as measured with non-resonant SHG in this
work. The specific ion effects on the orientation of the free O-H
group of the straddle type surface water can be more subtle. Since
SHG is not sensitive to the straddle type of surface water
molecules, further polarization dependent SFG-VS spectra measurement
and analysis have been conducted. The SFG-VS results on the specific
cation effects, especially the Na$^{+}$ and K$^{+}$ cations, shall
be reported elsewhere.\cite{FengRanRanManuscript}

\subsection{Complexities of the specific anion and cation effects on the interfacial water molecules}

With the data on the NaF, NaCl, NaBr, KF, KCl, and KBr solutions,
and with both the specific anion and cation effects established,
more detailed comparison between the anion and cation effects on the
interfacial water molecules can be explored.

In the bulk solution, the anion and the cation interactions with
water molecules can influence the macroscopic properties of
water,\cite{KunzCOCI2004,RecordJPCB5411} such as the surface
tension\cite{CollinsQRevBiophys} and the
viscosity.\cite{BinghamJPC885} The hydrated ions effects on the
hydrogen bonded networks in the bulk water are the key questions
people want to understand.\cite{KunzCOCI2004} For a long time, the
qualitative Hofmeister series first proposed more than a century ago
were used to explain the ion effects on forming and breaking of
hydrogen bonds in the aqueous
phase.\cite{CollinsQRevBiophys,CacaceQRevBiophys,FranKJCP507} It has
been believed that anions have more pronounced effects on the water
structure than the cations.\cite{CollinsQRevBiophys} In the mean
time, different cations can also influence the structure of water
molecules. For example, the different behaviors of Na$^{+}$ and
K$^{+}$ cations play important roles in many biochemical processes,
such as in the ion channels crossing the biological membranes.
Studies showed that these different behaviors were largely due to
their different hydration number and hydration free
energies.\cite{RutgersTFS2184,LisyJCP8555,LisyJCP024319,CarrilloTrippJCP118p7062}
Recently, using SFG vibrational spectroscopy, Cremer and co-workers
investigated the ion effects on the water molecules at the polymer
surfactant aqueous
interfaces.\cite{CremerJACS10522Hofmeister,CremerJACS12272IonEffect}
In these studies, significant anion effects and insignificant cation
effect were observed.

From the Hofmeister series in the bulk, the Na$^{+}$ cation was
traditionally called the `structure making' cation, while the
K$^{+}$ cation was considered `structure breaking' to some
extent.\cite{CollinsQRevBiophys} Using the mixed quantum mechanical
(QM) and molecular mechanical (MM) simulation, Tongraar \textit{et
al.} reported that the Na$^{+}$ cation can order the structure of
solvent molecules beyond the first hydration shell, while K$^{+}$
cannot go beyond the first hydration shell.\cite{TongraarJPCA10340}
So it was argued that the Na$^{+}$ be recognized as the
`structure-making', while K$^{+}$ be `structure-breaking'.
Mancinelli \textit{et al.} showed that water molecules are more
orientationally disordered around the K$^{+}$ ion than the Na$^{+}$
ion.\cite{Mancinelli13570} On the other hand, Saykally \textit{et
al.} used the X-ray absorption spectroscopy and showed that the
dissolved monovalent cations like Na$^{+}$ and K$^{+}$ did not have
significantly different effects on the nature of hydrogen bonds
around these ions.\cite{Saykally5301} Otma \textit{et al.} found,
using the femtosecond pump-probe spectroscopy, that the presence of
ions does not lead to an enhancement or a breakdown of the
hydrogen-bond network in liquid water.\cite{Omta347} There is still
no consensus regarding the ions effects on the hydrogen bond network
of the water molecules.

In the non-resonant SHG results, the KX series slightly increased
the orientational parameter D value of the interfacial water
molecules, while the NaX series slightly decreased the D value. The
larger its averaged D value, the more possibly tilted or disordered
the interfacial water molecules. This seems to suggest that the
Na$^{+}$ cation is `structure-making', while the K$^{+}$ is
`structure-breaking'. However, in the bulk Hofmeister series,
F$^{-}$ was usually considered `structure making', while the
Cl$^{-}$ and Br$^{-}$ `structure breaking'. If judged with the
change of D values, this seemed to work for the KX salt series, but
totally to the opposite for the NaX series, as shown in Figure
\ref{DThetaValueNaK}. Therefore, we can only conclude that the ion
effects at the electrolyte aqueous solution surfaces as observed in
the non-resonant SHG measurement may be more complex than the
Hofmeister series can qualitatively explain.

One example of the possible complexities is that further examination
of the $k$ (slope) values in the Table \ref{NumberPowerFitting}
suggested that the anions and cations may associate in some cases,
especially for the NaF aqueous solution surface. If there is no
anion and cation association at the interface, the $k$ values in the
Table \ref{NumberPowerFitting} should be able to factorize into ion
specific additive factors. This can be achieved easily for the $k$
values for the KF, KCl, KBr, NaBr and NaCl without considering the
$k$ value for the NaF. NaF stands out not only in this, but also in
its saturation concentration (less than 1M), which is far smaller
than the other five salts (all larger than 4.5M or more). It has
been long known that the F$^{-}$ is significantly different from the
larger Cl$^{-}$ and Br$^{-}$ ions in terms of interactions with the
water molecule.\cite{TobiasChemRev} However, there are huge
differences between the saturation solubility of the LiF (0.10M, or
0.27g LiF per 100g water), NaF(0.98M, or 4.13g NaF per 100g water)
and KF (15.4M, or 89.8g KF per 100g water) in water at
20$^{\circ}$C. In comparison, the saturation solubilities for the
chloride and bromide salts of these three cations are not so
different.\cite{CRCHandBook} These facts suggest that such
differences are most possibly due to the pairing or association of
the related anion and cation, since no single ion effect can explain
their differences. In the Table \ref{NumberPowerFitting}, the $k$
value of the NaF clearly stood out of the other five salts including
KF. Similar to the explanation for their bulk solubilities, this is
also most likely due to the Na$^{+}$ and F$^{-}$ ion pairing or
association, instead of the property of the F$^{-}$ anion alone.

Ion pairing is most likely to be found in aqueous solutions at high
bulk concentration. For example, a 5M monovalent salt solution has
less than six water molecules for each ion in the solution. In order
to fully solvate an ion in the solvation shell, usually six water
molecules are
needed.\cite{TongraarJPCA10340,LisyJCP024319,LisyJCP8555,CarrilloTrippJCP118p7062}
At such high bulk concentration as 5M the cations and anions are
very close to each other. However, our SHG data of the five salts
KF, KCl, KBr, NaBr and NaCl did not show the same ion pairing or
association effects as the NaF salt. Therefore, ion pairing or ion
association can not be simply defined by the average distance
between the cations and anions even at high bulk concentrations.

In the end, we have to realize that all the effects as measured with
the non-resonant SHG, i.e. the increase of the average thickness of
the interfacial water layer, as well as the shift of the averaged
orientational angle of the surface water molecules, are really
small. The increase of the thickness of the surface water layer is
most significant for the NaBr and KBr aqueous solutions. The
increase is only less than 30$\%$ for the 5M NaBr salt solution
(Figure \ref{WaterNumberNaK}). This increase is much smaller than
the more than 100$\%$ thickness increase value for the 2.0M (0.036
molar fraction) NaBr solution surface estimated using the relative
Raman, IR and SFG intensities in the
literature.\cite{AllenSodiumHalide2252} Increase of the surface
water layer thickness requires more ions to be solvated in the
surface water layer. Otherwise the surface concentration of the ions
could not keep up with the bulk ion concentration. Therefore, in
addition to the quantitative treatment in the non-resonant SHG, we
surmise that the slight increases of the average water surface layer
thickness as reported here are more acceptable. Furthermore, the
average orientational parameter D for the surface water molecules
also changed slightly for all the salt solutions studied (Figure
\ref{DThetaValueNaK}). All these suggested that the ion effects on
the surface water molecules were subtle effects. However, no matter
how weak these effects were, the presence and direct interaction of
the ions to the surface water molecules have been firmly
established. These conclusions may add vindications to the recent
interests on the specific ion effects at the air/water
surfaces.\cite{TobiasChemRev,WinterChemRev2006,JungwirthARPC}

In the experimental studies on the reactive and non-reactive
molecular beam scattering from the salty liquid solution surfaces,
such as from the glycerol salt solution surfaces, various cation and
anion effects have been
observed.\cite{NathansonJPCC112p3008,NathansonJPCC111p15043,NathansonJPCB110p4881}
Since molecular beam scattering would be difficult to be implemented
for the more volatile liquids than the glycerol, such as water, it
would be worthwhile to investigate the ion solvation and
interactions at these non-aqueous liquid surfaces with the nonlinear
optical techniques and to compare the results.

In summary, the specific ion effects on the surface water molecules
at the electrolyte aqueous solution surfaces can be more complex
than expected. Quantitative surface specific measurement such as the
non-resonant SHG can be employed to help establishment of the
detailed knowledge about these surfaces. Since the observed specific
cation and anion effects on the surface water molecules are in the
comparable magnitude, theoretical modeling and simulation efforts
are called upon to understand these unexpected cation effects for
the Na$^{+}$ and K$^{+}$ ions, as well as the unexpected anion
effects for the F$^{-}$ ions, which are usually considered to be
repelled from the aqueous solution surface.

\section{Conclusion}

In this report, using the procedures for sample control and
quantitative SHG data analysis, we studied the non-resonant SHG from
the water molecules at the aqueous solution surfaces of the
following monovalent salts: NaF, NaCl, NaBr, KF, KCl, and KBr. In
addition to the F$^{-}$, Cl$^{-}$ and Br$^{-}$ anion effects as
reported in our previous report, the specific Na$^{+}$ and K$^{+}$
cation effects on the interfacial water molecules are also clearly
and quantitatively identified for the first time.

Through quantitative polarization analysis of the non-resonant SHG
data, we found that the orientational order parameter D of the water
molecules at the interfacial region changed in different direction
for the NaX (X=F, Cl, or Br) series versus the KX series salts. The
average orientation angle of the dipole vector of the water
molecules at the NaX aqueous solution surfaces changed slightly to
the smaller tilt angle from the surface normal in comparison to that
of the neat air/water interface; while for the KX series, the
average orientation angle changed slightly to the larger tilt angle.
This seems to suggest that the Na$^{+}$ cation is more
``structure-making'', while the K$^{+}$ is more
``structure-breaking''. The relative ability for the different salts
to increase the average thickness of the interface water layer was
also quantitatively determined for the six salts. We found that
these values can be factorized into simple additive values for the
NaCl, NaBr, KF, KCl, and KBr salts except for the NaF. This
indicated that ion pairing or association effects for the Na$^{+}$
and F$^{-}$ ions should be much stronger than the cation and anion
in the other five salts.

Specific Na$^{+}$ and K$^{+}$ cation effects can also be observed in
the SFG-VS spectra of the interfacial water molecules at the NaF and
KF aqueous solution surfaces.\cite{FengRanRanManuscript} These
results on the specific cation and anion effects suggest that the
electrolyte aqueous solution surfaces can be more complex than the
currently prevalent theoretical and experimental understandings.
Since the observed specific cation and anion effects on the surface
water molecules are in comparable order of magnitudes, theoretical
modeling and simulation efforts are called upon to understand these
unexpected cation effects for the Na$^{+}$ and K$^{+}$ ions, as well
as the unexpected anion effects for the F$^{-}$ ions, which are
usually considered to be repelled from the aqueous solution surface.

\vspace{0.3cm}

\noindent\textbf{Acknowledgment:} HTB thanks the technical
assistance from De-sheng Zheng, Yan-yan Xu and An-an Liu. HFW thanks
the support by the Natural Science Foundation of China (NSFC,
No.20425309, No.20533070, No. 20773143) and the Ministry of Science
and Technology of China (MOST No. 2007CB815205). YG thanks the
support by the Natural Science Foundation of China (NSFC,
No.20673122).

\begin{figure}[h]
\begin{center}
\includegraphics[width=7cm]{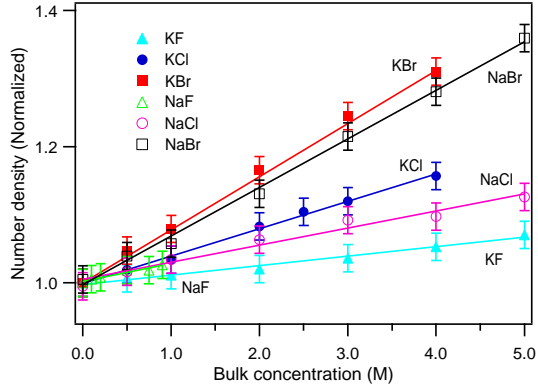}
\caption{The relative interfacial density of the water molecules at
different salt bulk concentration: KF (solid triangle), KCl(solid
circle), KBr (solid square), NaF (open triangle), NaCl(open circle)
and NaBr (open square). The data for KF, KCl and KBr are from Table
\ref{WaterFittingResults}. The data for NaF, NaCl and NaBr were the
same as in the previously published
work.\cite{BianPCCP}}\label{WaterNumberNaK}
\end{center}
\end{figure}
\begin{figure}[h]
\begin{center}
\includegraphics[width=7cm]{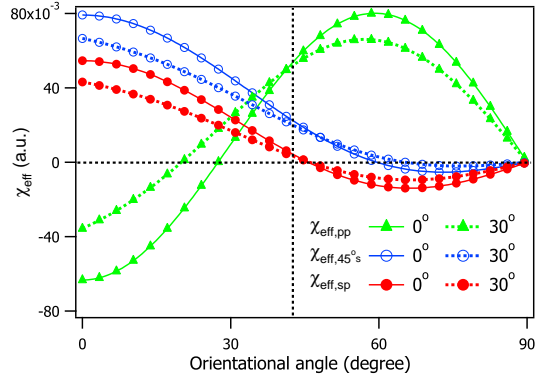}
\caption{The polarization dependent per molecule
hyperpolarizabilities ($\chi_{pp,0}$, $\chi_{45^{\circ}s,0}$, and
$\chi_{sp,0}$) are simulated with Eqs.\ref{xeff-chi} and
Eqs.\ref{ratio}. In calculating the Fresnel factors, the effective
surface dielectric constant was calculated according Zhuang
\textit{et al.}'s model.\cite{ShenZhuangPRB} The experimental
configuration is in the reflection geometry with
$\beta$=70$^{\circ}$. The values of R=$(\beta_{caa}+
\beta_{cbb})/(\beta_{aca}+ \beta_{bcb})$=0.69 and $(\beta_{aca}+
\beta_{bcb})$=1 are used. In the simulation, the angular
distribution function was assumed a Gaussian function
$f(\theta-\theta_{0})=\frac{1}{\sqrt{2\pi}\sigma}e^{-(\theta-\theta_{0})^2/2\sigma^2}$.
The values for $\delta$ distribution function ($\sigma=0^{\circ}$)
and for FWHM=30$^{\circ}$ ($\sigma$=12.7$^{\circ}$) are both
plotted. The vertical dash line is at the average orientational
angle ($\theta$) of the neat air/water interface which was
determined at 40.1$^{\circ}$. The horizontal dash line is drawn at
$\chi_{eff,0}$=0.} \label{XeffSimulation}
\end{center}
\end{figure}
\begin{figure}[h]
\begin{center}
\includegraphics[width=12cm]{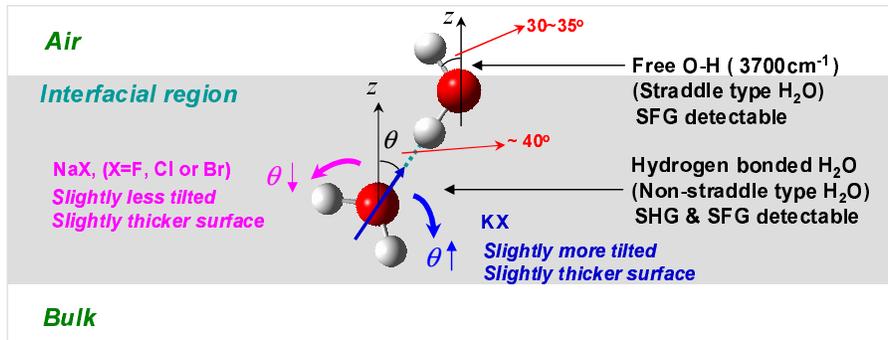}
\caption{Illustration of the two types of interfacial water
structure as well as the specific cation effects on the average
orientation of the non-straddle type surface water molecules. There
may exist different type of configurations for the non-straddle type
water molecules. Here only their averaged orientation of these
hydrogen bonded water molecules is representatively
illustrated.}\label{SurfaceWaterStructure}
\end{center}
\end{figure}

\clearpage
\begin{list}{}{\leftmargin 2cm \labelwidth 1.5cm \labelsep 0.5cm}

\item[\bf Table 1] The results for the $\chi _{zxx}/\chi _{xzx}$, $\chi _{zzz}/\chi
_{xzx}$, R, D, $\theta$ (assuming $\delta$ distribution) and N$_{s}$
for the neat water, KF, KCl and KBr salt solution interfaces.

\item[\bf Table 2] The relative ability for each salt to increase the thickness of the
interfacial water is determined by fitting the slopes in the Figure
\ref{WaterNumberNaK}. The six salts followed the following order:
KBr $>$ NaBr $>$ KCl $>$ NaCl $\sim$ NaF $>$ KF.

\end{list}

\clearpage

\begin{center}
\begin{table}
\caption[calculated results]{The results for the $\chi _{zxx}/\chi
_{xzx}$, $\chi _{zzz}/\chi _{xzx}$, R, D, $\theta$ (assuming
$\delta$ distribution) and N$_{s}$ for the neat water, KF, KCl and
KBr salt solution interfaces.}
\begin{tabular}{lcccccccccccccc}
\\
\hline
$KX$  & Concentration/M  &$\chi _{zxx}/\chi _{xzx}$   & $\chi _{zzz}/\chi _{xzx}$ & R & D & $\theta$    & $N_{s}$ \\
\hline
H$_{2}$O                 & 0.0 & 0.29$\pm$0.03  & 2.53$\pm$0.06 & 0.69$\pm$0.02 & 1.71$\pm$0.02 & 40.1$\pm$0.4 &1 \\
\hline
KF                    & 0.5 & 0.30$\pm$0.03  & 2.51$\pm$0.06  & 0.69$\pm$0.02 & 1.72$\pm$0.02 & 40.3$\pm$0.4  &1.01$\pm$0.02\\
                      & 1.0 & 0.30$\pm$0.03  & 2.57$\pm$0.06  & 0.69$\pm$0.02 & 1.72$\pm$0.02 & 40.3$\pm$0.4  &1.01$\pm$0.02\\
                      & 2.0 & 0.31$\pm$0.03  & 2.58$\pm$0.06  & 0.70$\pm$0.02 & 1.72$\pm$0.02 & 40.3$\pm$0.4  &1.02$\pm$0.02\\
                      & 3.0 & 0.31$\pm$0.03  & 2.58$\pm$0.06  & 0.70$\pm$0.02 & 1.72$\pm$0.02 & 40.3$\pm$0.4  &1.04$\pm$0.02\\
                      & 4.0 & 0.32$\pm$0.03  & 2.62$\pm$0.06  & 0.70$\pm$0.02 & 1.72$\pm$0.02 & 40.3$\pm$0.4  &1.05$\pm$0.02\\
                      & 5.0 & 0.33$\pm$0.03  & 2.60$\pm$0.06  & 0.71$\pm$0.02 & 1.72$\pm$0.02 & 40.3$\pm$0.4  &1.07$\pm$0.02\\
\hline
KCl                   & 0.5 & 0.29$\pm$0.03  & 2.52$\pm$0.06 & 0.69$\pm$0.02 & 1.71$\pm$0.02 & 40.1$\pm$0.4 &1.02$\pm$0.02\\
                      & 1.0 & 0.29$\pm$0.03  & 2.57$\pm$0.06 & 0.69$\pm$0.02 & 1.72$\pm$0.02 & 40.3$\pm$0.4 &1.03$\pm$0.02\\
                      & 2.0 & 0.28$\pm$0.03  & 2.53$\pm$0.06 & 0.68$\pm$0.02 & 1.71$\pm$0.02 & 40.1$\pm$0.4 &1.08$\pm$0.02\\
                      & 2.5 & 0.28$\pm$0.03  & 2.52$\pm$0.06 & 0.68$\pm$0.02 & 1.71$\pm$0.02 & 40.1$\pm$0.4 &1.10$\pm$0.02\\
                      & 3.0 & 0.29$\pm$0.03  & 2.56$\pm$0.06 & 0.69$\pm$0.02 & 1.72$\pm$0.02 & 40.3$\pm$0.4 &1.12$\pm$0.02\\
                      & 4.0 & 0.28$\pm$0.03  & 2.54$\pm$0.06 & 0.68$\pm$0.02 & 1.72$\pm$0.02 & 40.3$\pm$0.4 &1.16$\pm$0.02\\
\hline
KBr                   & 0.5 & 0.29$\pm$0.03  & 2.54$\pm$0.06 & 0.69$\pm$0.02 & 1.71$\pm$0.02 & 40.1$\pm$0.4 &1.05$\pm$0.02\\
                      & 1.0 & 0.28$\pm$0.03  & 2.54$\pm$0.06 & 0.68$\pm$0.02 & 1.72$\pm$0.02 & 40.3$\pm$0.4 &1.08$\pm$0.02\\
                      & 2.0 & 0.27$\pm$0.03  & 2.59$\pm$0.06 & 0.68$\pm$0.02 & 1.73$\pm$0.02 & 40.5$\pm$0.4 &1.17$\pm$0.02\\
                      & 3.0 & 0.27$\pm$0.03  & 2.61$\pm$0.06 & 0.68$\pm$0.02 & 1.73$\pm$0.02 & 40.5$\pm$0.4 &1.25$\pm$0.02\\
                      & 4.0 & 0.27$\pm$0.03  & 2.57$\pm$0.06 & 0.68$\pm$0.02 & 1.72$\pm$0.02 & 40.3$\pm$0.4 &1.31$\pm$0.02\\
\hline
\end{tabular}\label{WaterFittingResults}
\end{table}
\end{center}

\begin{center}
\begin{table}
\caption[fitting Ns power results]{The relative ability for each
salt to increase the thickness of the interfacial water is
determined by fitting the slopes in the Figure \ref{WaterNumberNaK}.
The six salts followed the following order: KBr $>$ NaBr $>$ KCl $>$
NaCl $\sim$ NaF $>$ KF.}
\begin{tabular}{lcccccccccccccc}
\\
\hline
Salts  & KBr  & NaBr   & KCl & NaCl & NaF & KF     \\
\hline
Slope (M$^{-1}$)  & 0.080$\pm$0.001 & 0.069$\pm$0.001  & 0.040$\pm$0.001 & 0.028$\pm$0.001 & 0.029$\pm$0.002 & 0.013$\pm$0.001  \\

\hline
\end{tabular}\label{NumberPowerFitting}
\end{table}
\end{center}

\clearpage
\begin{list}{}{\leftmargin 2cm \labelwidth 1.5cm \labelsep 0.5cm}

\item[\bf Fig. 1] Polarization and concentration dependent SHG (400nm) data:
a. left panel, KF, KCl and KBr solution surfaces; b. right panel,
NaF, NaCl and NaBr solution surfaces. These data were obtained by
fitting the s and p detection data measured for each concentration
and each salt solution. The procedure used was fully described
previously,\cite{ZhangWKJCP224713,BianPCCP} and it can ensure the
accuracy of the data for each polarization combination. The
45$^{\circ}$-in/s-out data need to be scaled up by the factor of
1.23 because of the different detection efficiency for the s and p
polarized SHG signal at 400nm in the detection setup in this
experiment. The data for NaF, NaCl and NaBr are identical to the
data in our previous article,\cite{BianPCCP} and the scale is
changed to be the same as the KF, KCl and KBr data for better visual
comparison.

\item[\bf Fig. 2] The orientational parameters D (top) and the apparent
orientational angles $\theta$ (bottom) for the KF (solid triangle),
KCl(solid circle), KBr (solid square), NaF (open triangle),
NaCl(open circle) and NaBr (open square) solution interfaces. The
data for KF, KCl and KBr are from Table \ref{WaterFittingResults}.
The data for NaF, NaCl and NaBr were from the Table 1 in the
previously published paper.\cite{BianPCCP} It is clear that the D
values for the neat water and low bulk concentration for the sodium
halides and the potassium halides data series were well within the
experimental error; while at higher concentrations the D values for
the two series went to different directions.

\item[\bf Fig. 3] The relative interfacial density of the water molecules at
different salt bulk concentration: KF (solid triangle), KCl(solid
circle), KBr (solid square), NaF (open triangle), NaCl(open circle)
and NaBr (open square). The data for KF, KCl and KBr are from Table
\ref{WaterFittingResults}. The data for NaF, NaCl and NaBr were the
same as in the previously published work.\cite{BianPCCP}

\item[\bf Fig. 4] The polarization dependent per molecule
hyperpolarizabilities ($\chi_{pp,0}$, $\chi_{45^{\circ}s,0}$, and
$\chi_{sp,0}$) are simulated with Eqs.\ref{xeff-chi} and
Eqs.\ref{ratio}. In calculating the Fresnel factors, the effective
surface dielectric constant was calculated according Zhuang
\textit{et al.}'s model.\cite{ShenZhuangPRB} The experimental
configuration is in the reflection geometry with
$\beta$=70$^{\circ}$. The values of R=$(\beta_{caa}+
\beta_{cbb})/(\beta_{aca}+ \beta_{bcb})$=0.69 and $(\beta_{aca}+
\beta_{bcb})$=1 are used. In the simulation, the angular
distribution function was assumed a Gaussian function
$f(\theta-\theta_{0})=\frac{1}{\sqrt{2\pi}\sigma}e^{-(\theta-\theta_{0})^2/2\sigma^2}$.
The values for $\delta$ distribution function ($\sigma=0^{\circ}$)
and for FWHM=30$^{\circ}$ ($\sigma$=12.7$^{\circ}$) are both
plotted. The vertical dash line is at the average orientational
angle ($\theta$) of the neat air/water interface which was
determined at 40.1$^{\circ}$. The horizontal dash line is drawn at
$\chi_{eff,0}$=0.

\item[\bf Fig. 5] Illustration of the two types of interfacial water structure
as well as the specific cation effects on the average orientation of
the non-straddle type surface water molecules. There may exist
different type of configurations for the non-straddle type water
molecules. Here only their averaged orientation of these hydrogen
bonded water molecules is representatively illustrated.

\end{list}

\clearpage

\end{document}